\documentclass[reprint,amsmath,prd,amssymb,aps,superscriptaddress]{revtex4-2}
\usepackage{graphicx}
\usepackage{dcolumn}
\usepackage{bm}
\usepackage{subfiles}
\usepackage{subcaption}
\usepackage{multirow}
\usepackage{soul}
\usepackage{csquotes}
\usepackage{dsfont}
\usepackage{gensymb}
\usepackage[table]{xcolor}
\usepackage[english]{babel}
\usepackage{bbm}
\usepackage{wrapfig}
\usepackage{slashed}
\usepackage{listings}
\usepackage{url}
\usepackage{hyperref}
\hypersetup{
  colorlinks   = true,    
  urlcolor     = blue,    
  linkcolor    = blue,    
  citecolor    = teal      
}

\begin{document}

\preprint{APS/123-QED}

\title{Systematic bias due to mismodelling precessing binary black hole ringdown}
\author{Cheng Foo}
\affiliation{
  Max Planck Institute for Gravitational Physics (Albert Einstein Institute), D-14476 Potsdam, Germany}
\affiliation{Institute for Theoretical Physics, ETH Zürich, Wolfgang-Pauli-Str. 27, 8093 Zürich, Switzerland}
\author{Eleanor Hamilton}
\affiliation{Departament de Física, Universitat de les Illes Balears,
IAC3 - IEEC, Crta. Valldemossa km 7.5, E-07122 Palma, Spain}
\affiliation{Physik-Institut, Universität Zürich, Winterthurerstrasse 190, 8057 Zürich, Switzerland}

\date{\today}

\begin{abstract}

Accurate waveform modelling is crucial for parameter estimation in gravitational wave astronomy, impacting our understanding of source properties and the testing of general relativity.
The precession of orbital and spin angular momenta in binary black hole (BBH) systems with misaligned spins presents a complex challenge for gravitational waveform modelling. 
Current precessing BBH waveform models employ a co-precessing frame, which precesses along with the binary. 
In this paper, we investigate a source of bias stemming from the mismodelling of ringdown frequency in the co-precessing frame.
We find that this mismodelling of the co-precessing frame ringdown frequency introduces systematic biases in parameter estimation, for high mass systems in particular, and in the Inspiral-Merger-Ringdown (IMR)-consistency test of general relativity. 
Employing the waveform model IMR\textsc{Phenom}XPHM, we conduct an IMR-consistency test using a Fisher matrix analysis across parameter space, as well as full injected signal parameter estimation studies. 
Our results show that this mismodelling particularly affects BBH systems with high mass ratios, high spin magnitudes, and highly inclined spins.  
These findings suggest inconsistencies for all waveform models which do not address this issue.

\end{abstract}

\maketitle

\section{\label{sec:1}Introduction}
Since the first detection of gravitational waves (GWs) in 2015, 90 GW events have been observed by the LIGO-Virgo-KAGRA Collaboration  (LVK), primarily originating from binary black hole (BBH) systems \cite{LIGO-O1-O2, LIGO-O3pt1, LIGOScientific:2020ibl, LIGO-O3pt2}. 
These systems, composed of two black holes orbiting around each other, emit GWs as they lose orbital energy, gradually spiralling towards each other before merging into a single, perturbed black hole. 
GWs contain information of the astrophysical systems that produced them, such as the masses and spins of the black holes, and can also be used to test our physical theory of general relativity (GR).

Parameter estimation, a process to extract properties of the source from GWs, relies on comparing accurate theoretical waveforms with GW data.
Systematic errors in theoretical waveforms can introduce biases in parameter estimation~\cite{Gupta:2024gun,    Dhani:2024jja, Hu:2022rjq}.
The increasing sensitivity of current detectors, in particular the LIGO, Virgo and KAGRA detectors \cite{lvknetwork, LIGOScientific:2014pky,VIRGO:2014yos,KAGRA}, and the construction of next generation detectors such as Cosmic Explorer \cite{Reitze:2019iox,Evans:2021gyd}, Einstein Telescope \cite{Punturo_2010} and LISA \cite{amaroseoane2017laser}, underscore the need for more precise waveform models. 
Current BBH waveform models used by the LVK can be split into three families, the phenomenological models ``\textsc{Phenom}" \cite{Hannam:2013oca, Husa:2015iqa, Khan:2015jqa, London:2017bcn, Khan:2018fmp, Khan:2019kot, Pratten:2020fqn, Pratten:2020ceb, Garcia-Quiros:2020qpx, Estelles:2020osj, Estelles:2020twz, Hamilton:2021pkf, Estelles:2021gvs, Thompson:2023ase}, the effective-one-body models ``SEOBNR" \cite{Taracchini:2012ig, Taracchini:2013rva, Pan:2013rra, Bohe:2016gbl, Cotesta:2018fcv, Ossokine:2020kjp, Pompili:2023tna, Ramos-Buades:2023ehm}, and the numerical relativity surrogate models ``NR\textsc{Surrogate}" \cite{Blackman:2015pia, Blackman:2017dfb, Blackman:2017pcm, Varma:2018mmi, Varma:2019csw, Williams:2019vub, Rifat:2019ltp, Islam:2021mha, Islam:2022laz}. 

GW signals from BBHs on quasi-circular orbits are characterised by the masses and spins of the binary.
Modelling efforts are complicated by eccentricity and spin effects.
In this paper, we are particularly interested in the effects of black hole spins with components in the orbital plane of the binary.
The orientation of spins in generic BBH systems can vary widely, depending on their formation process \cite{Gerosa:2018wbw,Kulkarni:2023nes}. 
When the spins of the two black holes in a binary are misaligned with the orbital angular momentum, both the orbital and spin angular momenta will precess about the total orbital angular momentum, a phenomenon known as (spin) precession \cite{Apostolatos:1994mx, Varma:2018rcg,Hannam:2013pra}. 
This results in intricate modulations of both the amplitude and phase of the GW signal, which are difficult to model \cite{Buonanno:2002fy, Schmidt:2010it}.
Waveform models for precessing BBHs usually employ a co-precessing frame, which precesses along with the binary. In this frame the waveform is well approximated by an aligned-spin waveform with a modified final spin \cite{Schmidt:2012rh}. 
In current models, this co-precessing frame is either defined by the orbital dynamics \cite{Pratten:2020ceb,Ramos-Buades:2023ehm,Ossokine:2020kjp}, or by the optimal emission direction~\cite{Hamilton:2021pkf,Thompson:2023ase}, which is the direction in which majority of the power is emitted. 
Current ``NR\textsc{Surrogate}" models instead use a co-orbital frame, which not only tracks the precession of the system using waveform quantities but also the instantaneous orbital phase, and interpolates numerical relativity waveforms in this frame \cite{Varma:2019csw}. 

In order to obtain the inertial frame waveform, a time- or frequency-dependent rotation is applied to the co-precessing waveform, e.g. see Refs. \cite{Estelles:2020osj, Estelles:2020twz, Varma:2018rcg, Hannam:2013pra, Hamilton:2021pkf, Ramos-Buades:2023ehm}. 
This procedure is called `twisting-up', and is performed with three time-/frequency-dependent Euler angles $\{\alpha, \beta, \gamma\}$.
Typically, these angles are chosen to describe the orientation of the co-precessing frame $z$ direction with respect to the total angular momentum $\bm{J}$. 

In this paper, we investigate a source of systematic bias in precessing BBH waveforms affecting the ringdown section of the waveform. 
Ringdown is the final stage of a GW signal from a BBH system, where the system can be modeled as a single perturbed black hole emitting GWs. These GWs are characterised by a set of complex frequencies, called quasinormal modes (QNMs), where the real parts are the ringdown frequencies \cite{Leaver:1985ax, Berti:2014fga}. 
Several works have shown complicated phenomenology in the ringdown for precessing BBH systems~\cite{OShaughnessy:2012iol, Hamilton:2023znn, Zhu:2023fnf}, which if mismodelled could lead to systematic bias in parameter estimation. 
The source of bias investigated here is due to employing the ringdown frequency derived from peturbation theory directly in the co-precessing frame, while the results from perturbation theory are only valid in the inertial $\bm{J}$-frame \cite{Hamilton:2021pkf,Hamilton:2023znn}. 
This mismodelling of co-precessing frame ringdown frequencies is present in many precessing models, such as IMR\textsc{Phenom}XPHM \cite{Pratten:2020ceb}, and SEOBNR\textsc{v4}PHM \cite{Ossokine:2020kjp}.
In this paper we will study IMR\textsc{Phenom}XPHM, one of the key models employed by the LVK in the analysis of detected GW signals.

Understanding this source of systematic bias is relevant for both the data analysis of high mass precessing BBH systems, as well as for tests of GR which rely on the ringdown \cite{Gupta:2024gun}. 
Several proposed tests of GR using GWs exist, and one such test is the Inspiral-Merger-Ringdown (IMR)-consistency test \cite{Ghosh:2017gfp, LIGOScientific:2019fpa}. The IMR-consistency test is based on GW data from a BBH merger. 
The fundamental concept underlying this test is that the mass and spin of the remnant black hole determined using two different segments of the GW signal, the inspiral (low-frequency) section and the ringdown (high-frequency) section, should be consistent~\cite{Hughes:2004vw}.
Assuming accurate parameter estimation, lack of consistency indicates a potential GR violation. 

Our analysis shows that mismodelling the co-precessing frame ringdown frequency leads to inaccuracies in GW data analysis. This is particularly so in highly precessing parameter space regions, especially at high values of mass ratio $q$, spin magnitude $\chi$ and spin misalignment with orbital angular momentum $\theta \sim 150\degree$. 
This will most strongly impact GW data analysis in two situations: first, when running parameter estimation on full signals from heavy mass systems, where the signal is dominated by the merger-ringdown; and second, when performing an IMR-consistency test. Our analysis shows one example of how mismodelling in the merger-ringdown portion of the signal can negatively impact parameter estimation and tests of GR.

We will first consider the impact of this source of systematic bias using Fisher analysis.
This enables us to study the effect of this source of bias in isolation.
We can therefore consider the impact not only on our ability to correctly measure the properties of the source from the GW signal, but also the consequent effect of these uncertainties on downstream analyses -- in particular tests of GR.
We will then consider a complete parameter estimation analysis.
This will give a more realistic picture of the impact on GW data analysis.
However, this analysis is less controlled and does not permit the investigation of a single source of waveform systematics.
We therefore consider only the effect on parameter estimation and leave considerations of tests of GR to future, improved models.

The structure of this work is as follows. In Sec. \ref{sec:precBBH}, we introduce the source of systematic bias under investigation in more detail. We then introduce the IMR-consistency test and the Fisher matrix formalism in Sec. \ref{sec:IMR} and Sec. \ref{sec:fisher}, as these are the main tools for our analysis. In Sec. \ref{sec:3}, we discuss the the methodology of our analysis, before presenting results across parameter space. Finally, in Sec. \ref{sec:4}, we perform signal injections and recovery using Bayesian analysis, to determine the effect of this bias on real parameter estimation.

\section{\label{sec:2}Preliminaries}
In general, a quasi-circular BBH system composed of two black holes with masses $m_1$ and $m_2$ can be described by 8 intrinsic parameters. These are the mass ratio of the binary $q=m_1/m_2$, the total mass of the system $M = m_1 + m_2$, and two black hole spins $\bm{S_1}$ and $\bm{S_2}$. We use the dimensionless black hole spins $\bm{\chi_1}$ and $\bm{\chi_2}$, where $\bm{\chi_i}= \bm{S_i}/m_i^2$. The spins can be decomposed into their components parallel and perpendicular to the direction of the Newtonian orbital angular momentum $\hat{\bm{L}}$, given by $\chi^\parallel_i = \bm{\chi}_i \cdot \hat{\bm{L}}$ and $\chi^\perp_i = \bm{\chi}_i - \chi^\parallel_i  \hat{\bm{L}}$ respectively. 

Dominant effects due to the spins of the black holes can be described by a combination of the spin components to two effective parameters $(\chi_\textrm{eff}, \chi_p)$. $\chi_\textrm{eff}$ describes dominant aligned-spin effects, and is given by \cite{Ajith:2009bn,Santamaria:2010yb}
\begin{equation}
    \chi_\text{eff} = \frac{m_1 \chi^\parallel_1+ m_2 \chi_2^\parallel}{m_1 + m_2}. 
   \end{equation}
On the other hand, $\chi_p$ describes dominant precession effects, parameterising spin perpendicular to the orbital angular momentum, and is given by 
\begin{equation} 
    \chi_{\rm p} =  \frac{S_{\rm p}}{m_1^2}, \label{eqn: chi_p}
\end{equation}
where $S_{\rm p} = \frac{1}{A_1} \text{max} \left( A_1 S_1^\perp, A_2 S_2^\perp \right)$, $A_1 = 2 + 3 m_2 /(2 m_1)$, and 
$A_2 = 2 + 3 m_1/(2 m_2)$ \cite{Schmidt:2014iyl}.

Since dominant spin effects are given by these two effective parameters, we can thus use a single-spin mapping to describe a two-spin system, which will give us the same phenomenology to leading order~\cite{Hamilton:2021pkf}. We will need signals of signal-to-noise ratios (SNRs) of 100 or more to make individual spin measurements, so we neglect two-spin effects for now \cite{Khan:2019kot}. 
For the rest of this paper, we consider only single-spin systems, where the more massive black hole has a spin, and the less massive black hole does not. In addition, we choose the convention where $m_1 \ge m_2$, and thus $q \ge 1$. The single spin on the larger black hole is described by two parameters, its magnitude $\chi$, and the angle it makes with $\bm{L}$, $\theta$. The parameter space we work with is  thus $(q, \chi, \theta)$.

\subsection{\label{sec:precBBH} Modelling ringdown in precessing BBH systems}

GWs are often decomposed into harmonics, which provide a useful basis for modelling the radiation pattern. However, the choice of harmonics depends on the nature of the source. GWs produced by the inspiral phase of a binary system are more conveniently modelled using spherical harmonics, while GWs produced by a black hole `ringing down' are best decomposed into spheroidal harmonics \cite{London:2017bcn,Garcia-Quiros:2020qpx,Berti:2009kk,Hamilton:2023znn}. 
It is common practice to decompose the full IMR signal into spherical harmonics \cite{London:2017bcn, London:2020uva}.

QNMs comprising the ringdown signal are uniquely determined by the mass and spin of the final black hole. The spin $s$-weighted QNM frequencies $\omega_{lmn}$, associated with projections onto spheroidal harmonics, are characterised by their angular number $l$ and azimuthal angle $m$. We consider only contributions from the fundamental QNMs ($n=0$)~\cite{Hamilton:2023znn}. For the rest of this paper the imaginary part $\textrm{Im}(\omega_{lm})$ is not relevant to us. When we refer to $\omega_{lm}$ from here onward, we are referring only to the real part $\textrm{Re}(\omega_{lm})$. In addition, we make the assumption that for the dominant $(2,|2|)$ multipole in the co-precessing frame, we can use the spheroidal characteristic QNM frequency to characterise the frequency in the spherical harmonic basis.  

As discussed in the Introduction, precessing waveform models employ a co-precessing frame model, which is then twisted up to an inertial frame waveform model using the Euler angles $(\alpha,\beta,\gamma)$. The angle $\beta$ is related to the opening angle of the precession cone during the inspiral \cite{Schmidt:2010it, OShaughnessy:2011pmr}. 
During the ringdown, $\beta$ is given by the orientation of the final spin relative to the optimal emission direction postmerger \cite{Hamilton:2023znn}. For further discussion on QNMs, see \cite{Berti:2009kk}, and for further discussion on the direction of emission for GWs from a perturbed black hole, see Sec. IX.B of \cite{Hamilton:2021pkf}. 

The rotation from the co-precessing frame to the inertial frame will introduce a shift in GW frequency, and since a rotation is still applied to the ringdown section of the waveform model, this means a shift in ringdown frequency \cite{Hamilton:2021pkf}. The shift in ringdown frequency from its co-precessing frame value $\omega^{\textrm{CP}}_{lm}$ to its inertial frame value $\omega^{\textrm{J}}_{lm}$ was derived in Ref. \cite{Hamilton:2023znn}, and is given by
\begin{equation}\label{eleanor_QNMfreq_eqn}
    \omega^{\textrm{J}}_{lm} =\omega^{\textrm{CP}}_{lm} + m(1-|\cos\beta_f|)\dot{\alpha},
\end{equation}
where $m$ is the QNM azimuthal number, $\dot{\alpha}$ is the time derivative of the Euler angle $\alpha$, and $\beta_f$ is the final ringdown value $\beta$ settles into after merger.

If the inertial frame ringdown frequency is applied in the co-precessing frame model, it will then be shifted away from the correct value when the model is twisted up to the inertial frame. The final model in the inertial frame will then have the wrong ringdown frequency. This is the case in some current and historical models, e.g. \cite{Pratten:2020ceb,Ossokine:2020kjp}, 
and has been corrected in more recent models \cite{Thompson:2023ase,Hamilton:2021pkf,Ramos-Buades:2023ehm}. In the rest of this paper, we investigate how this source of bias will affect data analysis using IMR\textsc{Phenom}XPHM. 

In the construction of the merger-ringdown section of IMR\textsc{Phenom}XPHM, the final mass and spin of the remnant $(M_f, \chi_f)$ are calculated using fits to numerical relativity (NR) and a correction to the final spin due to precessional effects \cite{Pratten:2020ceb}. The associated QNMs for these remnant parameters are then incorrectly employed in constructing the co-precessing frame ringdown waveform. 
As described above with Eq. (\ref{eleanor_QNMfreq_eqn}), the final $\omega_{lm}$ value (once rotated back to the inertial frame) will thus be wrong, having been shifted away from its true value. 
This inappropriate choice of co-precessing frame QNM frequency introduces systematic bias in the merger-ringdown section of the waveform model, which could lead to inaccurate parameter estimation results. 
In particular, data analysis of high mass precessing BBH systems will be affected, since more of the ringdown section of the signal will be in the LVK sensitivity band. 

It is important to understand where in parameter space these effects are most prominent, as we should be cautious with parameter estimation results in these regions. 
Tests of GR, including the IMR-consistency test, will also be affected. 
We anticipate this to be extreme regions of parameter space, with high values of $q$ and $\chi$, where precessional effects are more pronounced. 
Naively, we might expect that the effects of precession are maximised when the initial binary spins are fully in-plane, which would correspond to $\theta =90 \degree$.
However, this systematic bias is actually maximised at $\theta \approx 150 \degree$. 
This corresponds roughly to where $\cos(\beta_f) \approx 0$, as can be seen from Fig. 6 and the surrounding discussion in Ref. \cite{Hamilton:2023znn}. 
This would maximise the shift away from the true QNM ringdown frequency in Eq. (\ref{eleanor_QNMfreq_eqn}). 

Physically, this can be understood as follows. Ref. \cite{Hamilton:2021pkf} found, using NR data, that $\beta$ collapses through merger, rapidly dropping to a lower value. 
During inspiral, the precession cone (as measured by $\beta$) will be largest for large in-plane spins, where $\theta\sim90^\circ$.
For systems with higher mass ratios and close to anti-aligned-spins, $\beta$ can take values greater than $90^\circ$, although the precession cone is bounded by $90^\circ$ due to symmetry about the orbital plane.
For these systems, as $\beta$ collapses it moves closer to a value of $90^\circ$, and the precession cone actually increases, meaning that precession effects are maximal for these systems during merger and ringdown.
Thus, the optimal emission direction, which corresponds to the direction of the perturbation of the remnant \cite{Hamilton:2023znn}, will be close to orthogonal with final spin. 
The effect will thus be most pronounced for high values of $\theta$, excluding $\theta \approxeq 180\degree$, which will just stay close to anti-aligned with the final spin.

\subsection{IMR-Consistency Test}\label{sec:IMR}

Various tests of GR using GWs have been proposed \cite{LIGOScientific:2021sio}. 
One such test is the IMR-consistency test \cite{Ghosh:2017gfp, LIGOScientific:2019fpa}, which checks for consistency between two sections of the GW signal.

For the inspiral, which is the low-frequency part of the signal, initial masses and spins are first estimated using parameter estimation. These are then converted to remnant parameters using NR fits. The remnant parameters can also be similarly inferred from the post-merger, high-frequency part of the signal. 
These two sets of parameters are then compared. Assuming an accurate parameter estimation, we expect the two sets of estimates to be consistent with each other if GR is valid, and an inconsistency indicates a potential deviation from GR \cite{Ghosh:2017gfp, LIGOScientific:2019fpa}.

To have sufficient SNR in each of the sections of the waveform, the full signal is used, split into two at an appropriate cutoff frequency. This is in contrast to the original idea, which was to use part of the signal from the early inspiral, and part of the signal from the ringdown.
The frequency typically chosen is $f_\textrm{ISCO}$ \cite{Ghosh:2017gfp}, which corresponds to twice the orbital frequency of a test particle orbiting at the innermost stable circular orbit of the remnant black hole formed by the system \cite{Hofmann:2016yih, LIGOScientific:2019fpa, Ghosh:2017gfp}. The frequency $f_\textrm{ISCO}$ is not exactly the point where the inspiral regime moves to the merger regime, but it is a reasonable choice~\cite{Centrella:2010mx,Ghosh:2017gfp}. 

The mismodelling of the co-precessing ringdown frequency discussed in Sec. \ref{sec:precBBH} means that the ringdown waveform is inconsistent with the inspiral waveform. 
If such a model is used for an IMR-consistency test on precessing systems, this bias will lead to inconsistencies that are unphysical. 
To analyse this issue across parameter space, we assess the consistency of the inspiral and ringdown portions of the \textsc{IMRPhenomXPHM} model by extracting properties directly from the waveforms.
We determine whether the two estimates agree to within a given number of standard deviations.
To calculate this error, we use the Fisher-matrix formalism.

\subsection{Fisher matrix formalism}\label{sec:fisher}

When a GW passes through a detector, the recorded data $s(t)$ is
\begin{equation}
    s(t) = h(t) + n(t),
\end{equation}
where $h(t)$ is the GW strain, and $n(t)$ is the noise present during the detection. For the Fisher-matrix formalism, we assume that the noise is Gaussian and stationary \cite{Vallisneri:2007ev}. In data analysis, we look for the collection of parameters that characterise the GW signal $\theta = \{\theta_1,...,\theta_N\}$. For a true signal $h(t;\theta_{tr})$ inside GW data, described by real parameters $\theta_{tr}$, consider parameters $\theta^i_{tr}$ within the parameter set $\theta_{tr}$. Our estimates for $\theta^i_{tr}$ are given by maximum likelihood estimators $\hat{\theta}^i_{ML}$, and we can then express $\theta^i_{tr}$ as $\theta^i_{tr} = \hat{\theta}^i_{ML} + \Delta\theta^i$, where $\Delta \theta^i$ represents the errors for each of these parameters \cite{Maggiore:2007ulw}. 

The Fisher matrix is given by 
\begin{equation}\label{fisher}
    \Gamma_{ij} = \left( \frac{\partial h}{\partial \theta^i}  \middle|  \frac{\partial h}{\partial \theta^j} \right),
\end{equation}
evaluated at $\theta = \hat{\theta}_{ML}$. This object $\Gamma_{ij}$ is known as the \textit{Fisher information matrix} \cite{Cutler:1994ys}. It can be interpreted as the inverse of the covariance $\Sigma$ for our posterior probability distribution $p(\theta_{tr}|s)$ \cite{Vallisneri:2007ev}, 
\begin{equation}
    \langle \Delta\theta^i\Delta\theta^j \rangle = (\Gamma^{-1})^{ij} = \Sigma^{ij}.
\end{equation}
In essence, the Fisher information matrix quantifies the precision of our parameter estimation, and allows us to calculate the $1\sigma$ statistical errors in parameters $\theta_{tr}^i$. This $1\sigma$ value is simply given by the square root of the diagonal elements of the covariance matrix:
\begin{equation}
    \sigma^i = \sqrt{\Sigma^{ii}}=\sqrt{(\Gamma^{-1})^{ii}}
\end{equation}
The use of this formalism can thus provide us with a covariance matrix, assuming high SNR and stationary, Gaussian noise \cite{Vallisneri:2007ev}. 

For our specific analysis in Sec. \ref{sec:3}, we will analyse a system with a three-dimensional parameter space described by $(q,\chi,\theta)$. The relevant Fisher matrix for such a system can be written as follows:
\begin{equation}\label{eqn:fisher3dim}
        \Gamma^{(3)} = 
        \begin{pmatrix}
            \Gamma_{qq} & \Gamma_{q\chi} & \Gamma_{q\theta} \\
            \Gamma_{\chi q} & \Gamma_{\chi \chi} & \Gamma_{\chi \theta}\\
            \Gamma_{\theta q} & \Gamma_{\theta \chi} & \Gamma_{\theta\theta}
        \end{pmatrix}
\end{equation}
We need derivatives of the waveform to calculate each of these components, as in Eq. (\ref{fisher}), and we solve for these numerically. 

The covariance matrix for our inspiral three parameter system is $(\Gamma^{(3)})^{-1} = \Sigma^{(3)}$. For our analysis, we will need to convert this into the covariance matrix for our remnant parameters, 
\begin{equation}\label{eq:remparamcovariancematrix}
    \Tilde{\Sigma} = \begin{pmatrix}
        \Tilde{\Sigma}_{M_f M_f} & \Tilde{\Sigma}_{M_f \chi_f} \\
        \Tilde{\Sigma}_{\chi_f M_f} & \Tilde{\Sigma}_{\chi_f \chi_f}
    \end{pmatrix}.
\end{equation}
Following the method set out in Ref. \cite{Bhat:2022amc}, the elements of the matrix above can be calculated as follows:
\begin{align}
    \Tilde{\Sigma}_{M_f M_f} &= \delta M_f^2 = \sum_{i,j }\left( \frac{\partial M_f}{\partial \theta ^i}\right) \left( \frac{\partial M_f}{\partial \theta ^j}\right) \Sigma^{(3)}_{ij}, \\
    \Tilde{\Sigma}_{\chi_f \chi_f} &= \delta \chi_f^2 = \sum_{i,j }\left( \frac{\partial \chi_f}{\partial \theta ^i}\right) \left( \frac{\partial \chi_f}{\partial \theta ^j}\right) \Sigma^{(3)}_{ij}, \\
    \Tilde{\Sigma}_{M_f \chi_f} &= \Tilde{\Sigma}_{\chi_f M_f }  = \sum_{i,j }\left( \frac{\partial M_f}{\partial \theta ^i}\right) \left( \frac{\partial \chi_f}{\partial \theta ^j}\right) \Sigma^{(3)}_{ij}.
\end{align}

The covariance matrix can be represented as an error ellipse in plots. To plot our error ellipses in the $M_f - \chi_f$ plane, we need the semi-major axis $a$, the semi-minor axis $b$, and the angle $\theta_\textrm{ellipse}$, which is the counter-clockwise angle relative to the horizontal axis $M_f$. In order to obtain $a$ and $b$, we need to calculate the eigenvalues of our covariance matrix, $\Lambda_\pm$. For a $1\sigma$ error ellipse, the semi-major axis $a$ and the semi-minor axis $b$ are related to the eigenvalues as follows \cite{Bhat:2022amc}:
\begin{align}
    a &= \sqrt{\Lambda_+} & b &= \sqrt{\Lambda_-}.
\end{align}
The angle $\theta_\textrm{ellipse}$ can be calculated from the final parameter covariance matrix \cite{Bhat:2022amc}, 
\begin{equation}
    \theta_\textrm{ellipse} \approx -\frac{1}{2} \arctan \left( \frac{2\Tilde{\Sigma}_{M_f \chi_f}}{  \Tilde{\Sigma}_{M_f M_f} \Tilde{\Sigma}_{\chi_f \chi_f} } \right).
\end{equation}

\section{\label{sec:3}IMR-Consistency Test Study}

In this Section, we conduct an `IMR-consistency test' on waveforms generated by the IMR\textsc{Phenom}XPHM model. Our goal here is to investigate the source of systematic bias described in Sec. \ref{sec:precBBH}, and to determine the region of parameter space where it will affect data analysis.

We begin by generating a waveform instance for a given binary configuration with specific input inspiral parameters using \texttt{PyCBC} \cite{pycbcv2.0.5}. 
We then split the waveform instance into two sections at an appropriate cutoff frequency, which we choose to be $f_{\textrm{ISCO}}$~\cite{Bardeen:1972fi, Favata:2021vhw}. For each of our two sections of the waveform, we employ distinct methods to estimate the parameters of the remnant black hole $(M_f, \chi_f)$, followed by calculating the error ellipses using the Fisher matrix formalism. 
We choose to employ the Fisher matrix formalism as we want to conduct our analysis across parameter space, and thus need a computationally efficient method. 

If the ellipses for a particular confidence interval do not overlap, it indicates that the final parameters estimated from the two sections of our waveform are incompatible with each other up to that confidence level. In accordance with the IMR-consistency test criteria, this result implies a `violation of GR'.
Since the model is intended to faithfully represent a GW signal as predicted by GR, this violation is non-physical and comes from inaccuracies in the waveform model.

The Fisher analysis enables us to identify the regions of parameter space where model inaccuracies will lead to this erroneous conclusion, due to systematic biases.
We explore the parameter space described by $(q, \chi, \theta)$. We conduct our analysis across this parameter space for waveforms with specific SNR values, and look for predicted `violations of GR' at $1\sigma$, $2\sigma$ and $3\sigma$ confidence levels.

\subsection{\label{sec:calculatingremnantfisher}Calculating Remnant Parameters}
\subsubsection{Inspiral}\label{inspiralremnantparam}

The inspiral part of the waveform is characterised by the input intrinsic parameters. We can calculate the values of $(M_f,\chi_f)$ predicted for a binary with these input parameters using fits obtained from NR simulations.

We follow the method that IMR\textsc{Phenom}XPHM uses to model the final state of a precessing system given the input inspiral parameters (c.f. Section IV.D of \cite{Pratten:2020ceb}). 
For inspiral parameters $(q,\chi, \theta)$, we first find the final state $(M_f^\parallel, \chi_f^\parallel)$ of the corresponding aligned-spin system parameterised by $(q,\chi\cos\theta,0)$, using the NR fits from \cite{Jimenez-Forteza:2016oae}. 
These fits are implemented using code from package \texttt{llondon6/positive} \cite{llondon_positive}.
We then employ approximations to obtain the final state of the precessing system~\cite{Pratten:2020ceb}.

For the final mass, several NR fit studies have shown that there is only a weak dependence on precession \cite{finalmassandspin1,Varma:2018aht,Siemonsen:2017yux}. Thus, we follow IMR\textsc{Phenom}XPHM and its predecessors in setting $M_f = M_f^\parallel$.

For the final spin, we cannot ignore precession dependence \cite{Pratten:2020ceb}. 
The final spin is calculated from the aligned-system final spin $\chi^\parallel_f$ using
\begin{equation}
    \chi_f=\textrm{sgn}\ (\chi_f^\parallel)\sqrt{\left(\chi_p\ \frac{m_1^2}{(m_1+m_2)^2}\right)^2+\left(\chi^\parallel_f\right)^2}.
\end{equation}
This equation is adapted from \cite{pv2notes,Hannam:2013oca}. For single-spin systems with the spin on the larger black hole, $\chi_p = \chi \sin \theta$ is the in-plane spin component. 
This equation is then equivalent to
\begin{equation}
    \chi_f = \textrm{sgn}\ (\chi_f^\parallel)\sqrt{\left(\chi\sin\theta\frac{q^2}{q^2 + 2q + 1}\right)^2+\left(\chi^\parallel_f\right)^2}.
\end{equation}
We will denote these final mass and spin values calculated from the inspiral section as $(M^{\textrm{insp}}_f, \chi^{\textrm{insp}}_f)$.

\subsubsection{Merger-Ringdown} \label{ringdownparamestimate}

Now we compute the remnant parameters from the post-merger section of our waveform, which we denote $(M^{\textrm{merg}}_f, \chi^{\textrm{merg}}_f)$. This computation is more complicated than the inspiral equivalent above -- since the IMR\textsc{Phenom}XPHM waveforms are generated with inspiral input parameters, 
we need to extract the parameters directly from the generated merger-ringdown waveform. As discussed above, NR studies have shown the final mass value has a weak dependence on precession. We thus take $M^{\textrm{merg}}_f = M^{\textrm{insp}}_f$ and assume that the source of bias outlined in Sec~\ref{sec:precBBH} will affect only $\chi_f$. This assumption allows us to estimate the final spin $\chi_f^\mathrm{merg}$ from the merger-ringdown section.

To accomplish this, we analyse the QNM spectrum of the final black hole. In particular, we want to extract the characteristic ringdown frequency $\omega_{22}$ of the $(2,|2|)$ mode. Using QNM data tables \cite{Berti:2009kk}, which list $\omega_{lm}$ values for corresponding black hole masses and spins, we can then compute $\chi_f$.

We generally expect a characteristic dip in the phase derivative of the aligned-spin waveform at frequency $f_0$. This dip frequency coincides up to a small error with the ringdown frequency of the final black hole for an aligned-spin system, $\omega_{22}$ \cite{Husa:2015iqa, Khan:2015jqa, Garcia-Quiros:2020qpx}. 
Across parameter space, the behaviour of the phase derivative of the $(2,|2|)$ mode is relatively well behaved for aligned-spin systems, and for precessing systems in the co-precessing frame. However, trying to search for this dip in the inertial frame phase derivative of a precessing waveform is not trivial, due to the mixing of spherical multipoles caused by the twisting up procedure~\cite{Hamilton:2023znn}. 
As precession effects grow, the behaviour of the phase derivative is less well-behaved. 

\begin{figure}[t]
    \centering
    \begin{subfigure}[t]{0.5\textwidth} 
        \centering
        \includegraphics[width = 1.0\textwidth]{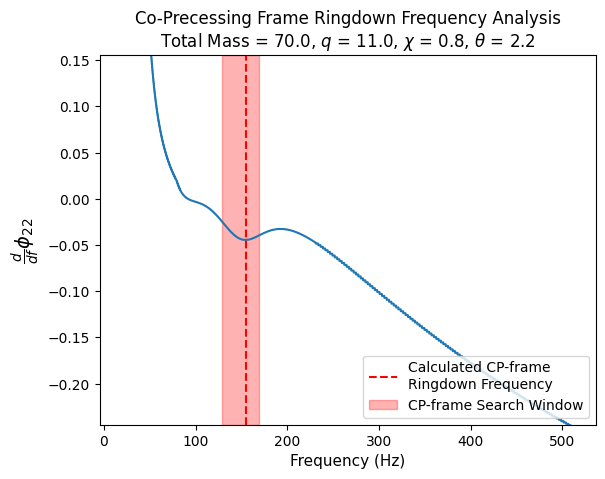}
    \end{subfigure}
    \begin{subfigure}[t]{0.5\textwidth} 
        \centering
        \includegraphics[width = 1.0\textwidth]{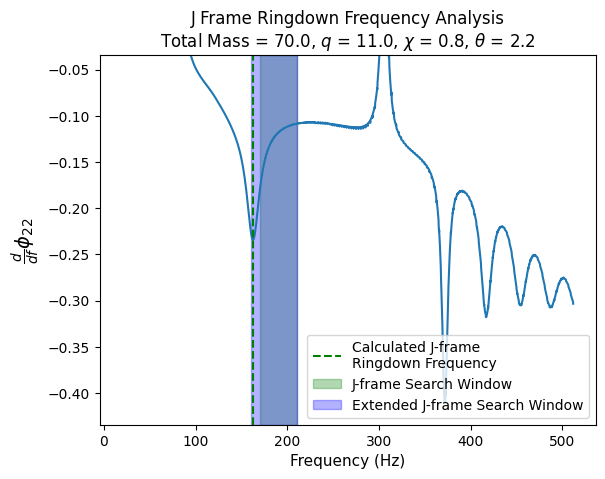}        
    \end{subfigure}
    \caption{Two figures demonstrating our search for $\omega_{22}$, one in the co-precessing frame (above), and one in the inertial frame (below). The dotted lines show the determined $\omega_{22}$.}
    \label{fig:dipsearch}
\end{figure}

We first find an estimate of $\omega_{22}$ using the predicted remnant mass and spin values calculated in Sec. \ref{inspiralremnantparam}, 
and then computationally search for the characteristic dip in $d\phi_{22}/df$ in a window around this estimate. An example of such a search, in both the co-precessing frame and the inertial frame, can be seen in Fig. \ref{fig:dipsearch}. Our results for $\omega_{22}$ are relatively noisy for high $q$, $\chi$ and $\theta$ values. 
In order to analyse general trends across parameter space, we generate $\omega_{22}(\theta)$ values across $\theta$ parameter space for each $\{q,\chi\}$ used in the analysis, and perform a cubic fit through these values. The QNM frequency we use is then the value of this fit at our input $\theta$.

To convert this QNM frequency into final spin, we create a spline from the QNM data tables \cite{Berti:2009kk}, inputting $\omega_{22}$ and $M^{\textrm{merg}}_f$ to obtain $|\chi^{\textrm{merg}}_f|$. We follow IMR\textsc{Phenom}XPHM in setting $\textrm{sgn}(\chi_f) = 1$ for prograde QNM frequencies, and $\textrm{sgn}(\chi_f) = -1$ for retrograde QNM frequencies. 

\subsubsection{Final parameter difference} \label{finalpardiffsection}

\begin{figure*}
    \centering
    \includegraphics[width = 1.0\textwidth]{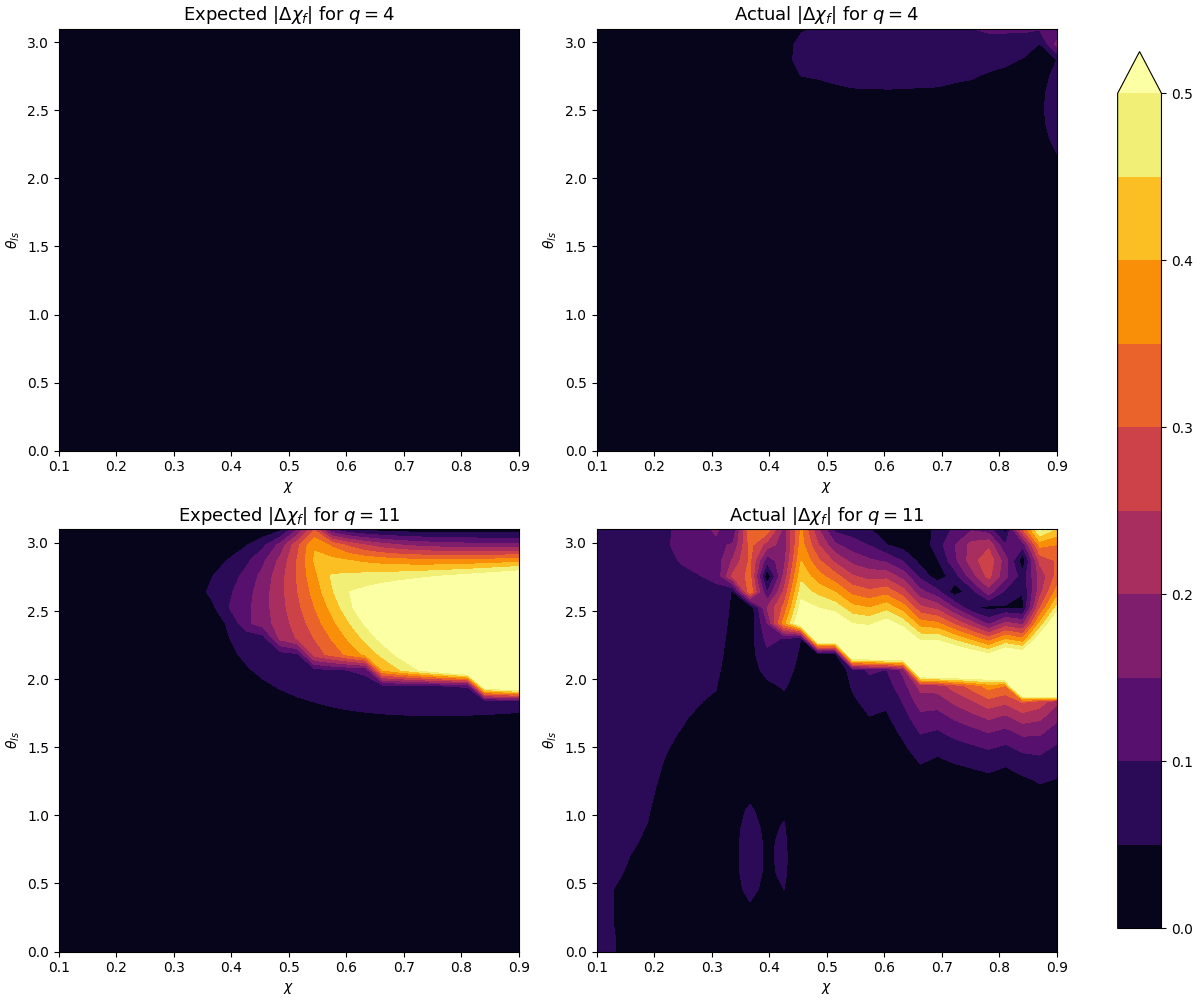}
    \caption{Figures on the left show the expected final spin difference $|\Delta\chi_f^\mathrm{expect}|$ between inspiral and merger, resulting from the shift of QNM frequencies identified in Eq. (\ref{eleanor_QNMfreq_eqn}). Figures on the right show the final spin difference $|\Delta\chi_f^\mathrm{model}|$ found between inspiral and ringdown from IMR\textsc{Phenom}XPHM waveforms. All systems were generated with a total mass of 70$M_\odot$.}
    \label{fig:FinalParDiff}
\end{figure*}

As a sanity check, we want to make sure that the discrepancy between $\chi^{\textrm{merg}}_f$ and $\chi^{\textrm{insp}}_f$ is entirely due to the mismodelling of the co-precessing ringdown frequency discussed in Sec. \ref{sec:precBBH}. To accomplish this, we need to know what the expected shift in final spin $|\Delta\chi_f^\textrm{expect}|=|\chi^{\textrm{shifted}}_f - \chi^{\textrm{insp}}_f|$ should be, where $\chi^{\textrm{shifted}}_f$ is the final spin after the shift of ringdown frequency away from its true value. 
We calculate the QNM frequency $\omega_{22}^\mathrm{insp}$ associated with our inspiral remnant parameters $(M^{\textrm{insp}}_f, \chi^{\textrm{insp}}_f)$, and perform the expected shift using Eq. (\ref{eleanor_QNMfreq_eqn}), taking the Euler angles generated using the package \texttt{Cyberface/gw-phenom} \cite{gw-phenom}. 
These are the angles derived from an application of multiple scale analysis (MSA) \cite{Chatziioannou:2017tdw}, as utilised by IMR\textsc{Phenom}XPHM.
Using the shifted QNM frequency and $M^{\textrm{insp}}_f$, we then calculate $\chi^{\textrm{shifted}}_f$ from QNM data table splines as above \cite{Berti:2009kk}. 
Finally, we compute
\begin{align} 
   |\Delta\chi_f^\mathrm{model}| &= |\chi_f^\mathrm{merg} - \chi_f^\mathrm{insp}|, \\
   |\Delta\chi_f^\mathrm{expect}| &= |\chi_f^\mathrm{shifted} - \chi_f^\mathrm{insp}|,
\end{align}
across the $(\chi, \theta)$ parameter space for discrete choices of $q$.

As can clearly be seen from Fig. \ref{fig:FinalParDiff}, the similarities between the expected and actual final spin difference show that the discrepancy in final spin estimates is mainly due to the systematic bias from $\omega_{22}$ mismodelling, rather than other systematics. Note that we are not yet considering any statistical errors in our estimates of the remnant parameters -- that is something we will look at next in Sec. \ref{subsecfish}.

The results here agree with our expectations that the difference in estimates will be more pronounced for extreme regions in parameter space. For $q=4$, the two spin estimates agree generally, while for $q = 11$, 
there is a dramatic increase in $|\Delta\chi_f|$ at high $\chi, \theta$ values. This is what we expect -- as precession effects become more relevant, the shift in the ringdown frequency should increase, leading to a greater difference.

\subsection{\label{subsecfish}Computing Errors in Remnant Parameters}
For each $ (q,\chi,\theta)$ input, we calculate the uncertainties in measurements of $(M_f,\chi_f)$ from both parts of the waveform for a signal at a given SNR. We do this using the Fisher matrix formalism detailed in Sec. \ref{sec:fisher}, calculating the 3-dimensional Fisher matrix Eq. (\ref{eqn:fisher3dim}) for both the inspiral and for the merger-ringdown. We use numerical derivative methods to calculate the derivatives of the final parameters with respect to the input parameters. This is trivial for the inspiral, as the input parameters are simply the inspiral parameters. For the merger, this is more complicated. For example, for calculating the derivative of $M^{\textrm{merg}}_f$ with respect to $q$, we repeat the entire analysis detailed in Section \ref{ringdownparamestimate} for $q+\delta q$, and for $q-\delta q$, and perform a numerical derivative. 

Having obtained the 2-dimensional covariance matrix of the remnant parameters, following the analysis in Sec. \ref{sec:fisher} to obtain Eq. (\ref{eq:remparamcovariancematrix}), we can then compute the error ellipses for the final mass and spin. Putting all above results together, we perform an IMR-consistency test on the waveform model IMR\textsc{Phenom}XPHM across a large region of parameter space.

\begin{figure*}[t]
  \centering
  \begin{subfigure}[t]{0.33\textwidth}
      \centering
      \includegraphics[width = 1.0\textwidth]{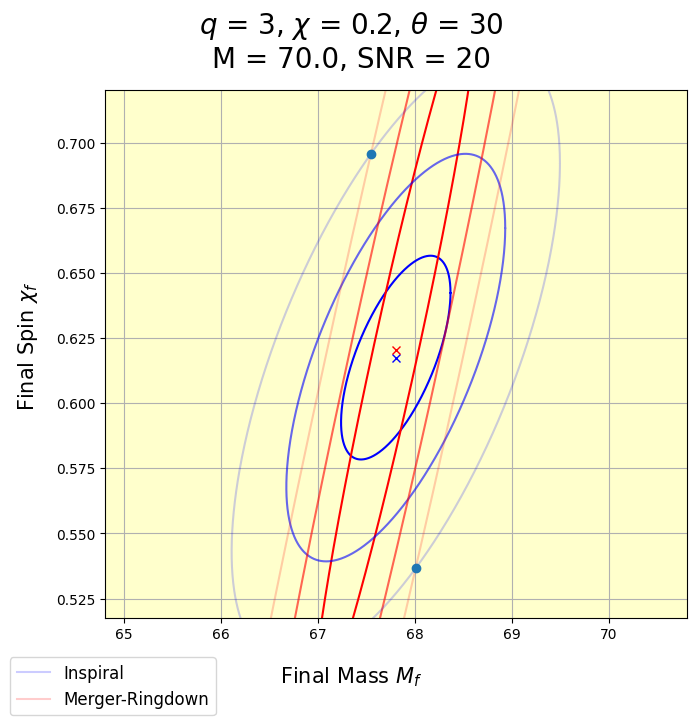}
      \caption{Consistent at $1,2$ and $ 3\sigma$.}
  \end{subfigure}\hfill
  \begin{subfigure}[t]{0.33\textwidth}
      \centering
      \includegraphics[width = 1.0\textwidth]{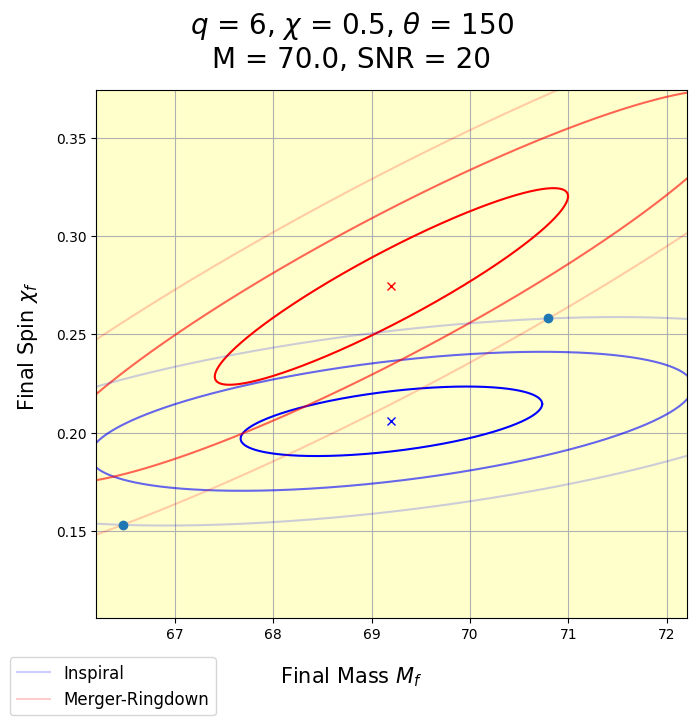}
      \captionsetup{width=0.9\linewidth}
      \caption{Consistent at $2$ and $3\sigma$, inconsistent at $1\sigma$.}
      \label{fig: fisherexample-no-violation}
  \end{subfigure}\hfill
  \begin{subfigure}[t]{0.33\textwidth} 
      \centering
      \includegraphics[width = 1.0\textwidth]{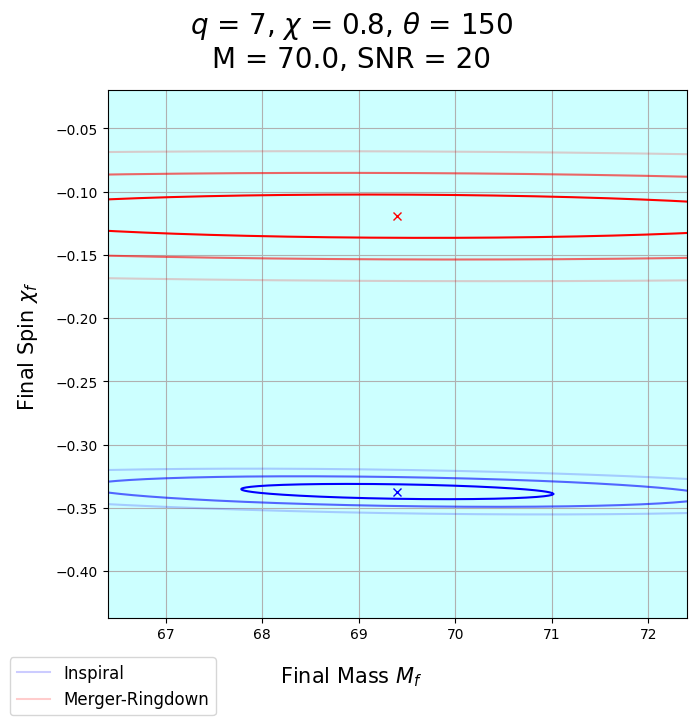}
      \captionsetup{width=0.9\linewidth}
      \caption{Inconsistent at $1,2$ and $ 3\sigma$.}
      \label{fig: fisherexample-violation}
  \end{subfigure}
  \caption{Fisher formalism IMR-consistency analysis results for 1, 2 and 3$\sigma$ confidence levels are shown in the figure. The input inspiral parameters are displayed at the top of each figure. The recovered parameters $(M_f, \chi_f)$ from the inspiral and merger-ringdown of the waveform are marked with `x' points, and the error ellipses, calculated using the Fisher matrix method, are plotted centered on these points. Consistency between the inspiral and merger-ringdown spin estimates at a specific confidence level occurs if the the error ellipses generated by the two sections overlap. Results demonstrating consistency at $3\sigma$ level are highlighted with a yellow background, while inconsistency is denoted by a blue background.}
  \label{fig:FisherResultsExample}
\end{figure*}

\subsection{\label{sec:imrresults}Results of analysis across parameter space}

Our findings reveal that the model incorrectly predicts violations of GR in more extreme regions of parameter space (i.e., high $q$, $\chi$, and $\theta \approx 150\degree$), inline with our expectations described in Sec. \ref{sec:precBBH}. We also see fewer predicted violations at lower SNR. 
This makes sense, as we expect our error ellipses to scale with $1/\sqrt{SNR}$ \cite{Maggiore:2007ulw}. If our error ellipses are larger, systematic bias effects are less pronounced. This also means that as SNR increases, we see violations across a larger region of parameter space, as for larger SNRs our error ellipses should be smaller.

Fig. \ref{fig:FisherResultsExample} illustrates the results of our analysis for three specific cases, at different points from widely separated regions in parameter space to demonstrate the observed parameter space trend. We investigate at confidence levels of 1$\sigma$, 2$\sigma$, and 3$\sigma$, plotting $x\sigma$ error ellipses centered around the values estimated in Section \ref{sec:calculatingremnantfisher}. In a standard IMR-consistency test, non-overlapping $x\sigma$ error ellipses indicates a potential deviation from GR at $x\sigma$ level. In our analysis here, such non-overlap would constitute an inconsistency in the model for the given input parameters, warranting caution in these specific parameter space regions.

Our results follow our expectations as outlined in Sec. \ref{sec:precBBH}. The leftmost figure shows results for a weakly precessing system with low values of $(q,\chi,\theta)$, where even the $1\sigma$ error ellipses overlap. Therefore, the model is consistent for this system. However, this is not the case for the other two configurations. 
The middle figure shows a system with $(q=6, \chi =0.5, \theta=150\degree)$.
Here, the model is consistent at  $2$ and $3\sigma$ level, but is inconsistent at $1\sigma$. 
The rightmost figure shows the most strongly precessing system with $(q=7, \chi =0.8, \theta=150\degree)$. Here we see that the model predicts a violation at all $\sigma$ levels. 
We can clearly see that as we move towards more highly precessing systems, the deviation between the final spin estimate calculated using the two methods increases from $\mathcal{O}(10^{-3})$ (left hand panel) to $\mathcal{O}(10^{-1})$ (right hand panel). Therefore, although the fractional uncertainty estimate remains constant, the error ellipses are much more strongly separated in the case of a highly precessing system.

Having examined a few specific cases, we now aim to systematically explore trends across parameter space. To this end, we perform a similar analysis to the one described above for $28\times 28 \times 28 = 21,952$ points in parameter space. 
This analysis is conducted for a given total mass, SNR, and $\sigma$ level. Although most of the violation points appear in the region where we expect them, we still observe some scattered throughout the rest of parameter space. To identify general trends, we apply a noise reduction technique to our data. 

\begin{figure*}
  \centering
  \begin{subfigure}[t]{1\textwidth} 
      \centering
      \includegraphics[width = 0.8\textwidth]{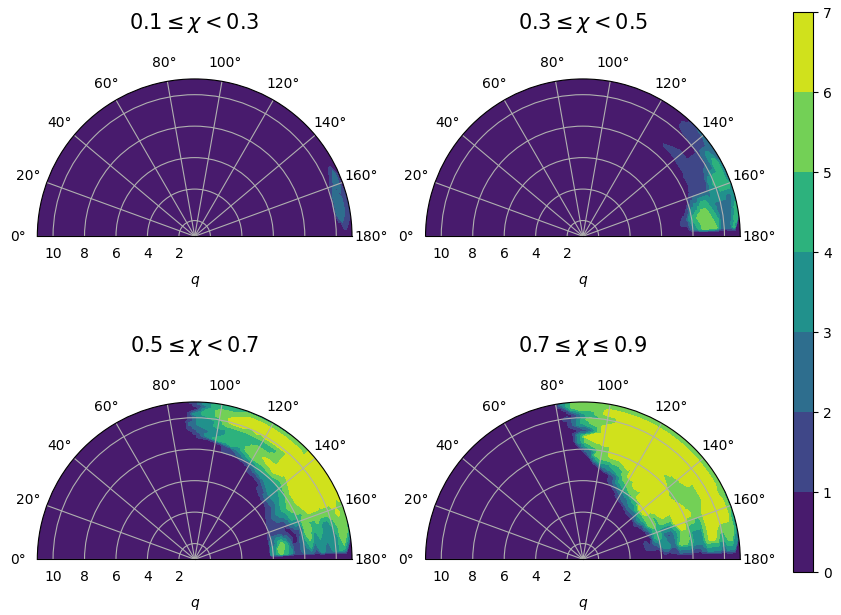}
      \captionsetup{width=1.0\linewidth}
      \caption{IMR-consistency test violation points at 3$\sigma$, for systems with total mass $M=70M_\odot$ and SNR $=20$.}
      \label{fig:2,3sigma}
  \end{subfigure}
  \begin{subfigure}[b]{1\textwidth} 
      \includegraphics[width = 0.8\textwidth]{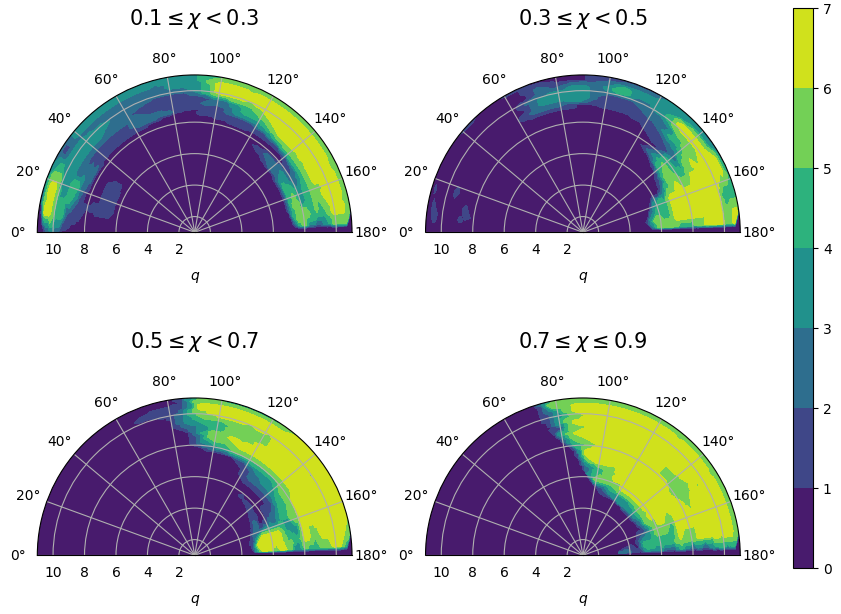}
      \captionsetup{width=1.0\linewidth}
      \caption{IMR-consistency test violation points at 3$\sigma$, for systems with total mass $M=70M_\odot$ and SNR $=50$.}
      \label{fig:5,3sigma}
  \end{subfigure}
  \caption{These plots illustrate the results of an IMR-consistency test analysis conducted across parameter space, for $28\times28\times28$ points. The results are split into four bins depending on the $\chi$ input value. The radial axis shows $q$, and the angular axis shows $\theta$. In each $\chi$ bin, there are 7 possible $\chi$ points for each $(q,\theta)$. The plots here show the number of violations across $\chi$ for each $(q,\theta)$, in each bin, with the colour bar showing this number. We clearly see more violations for higher $\chi$, and clustered around high $q$ and $\theta$. From the difference between plots (a) and (b), we see the effect that SNR has on this analysis.}    
  \label{fig:paramspace}
\end{figure*}

The results, as seen in Fig. \ref{fig:paramspace}, show violations in the high $(q,\chi,\theta)$ regions of parameter space as expected. We see that as each of the parameters gets closer to the region of expected violation (high $q$ and $\chi$, $\theta$ approaching $150\degree$), there is an increased clustering of violation points.
Comparing Fig. \ref{fig:2,3sigma} to Fig. \ref{fig:5,3sigma}, we see that an increase in SNR correlates with a broader violation region. 

We also performed this analysis for varying values of SNR, and at 1$\sigma$, 2$\sigma$, and 3$\sigma$. 
In general, the region of parameter space in which we see inconsistencies at a given $\sigma$ increases for higher SNR.
This result aligns with the expectation that increased uncertainty in our estimates, reflected by larger error ellipses, facilitates overlooking systematic biases. This also means that as our detectors become more sensitive, and we begin detecting highly precessing systems, this bias will be more significant.

\section{\label{sec:4}Parameter Estimation}
\begin{figure*}[t]
  \centering
  \begin{subfigure}[t]{1.0\textwidth}
      \centering
      \includegraphics[width = 0.93\textwidth]{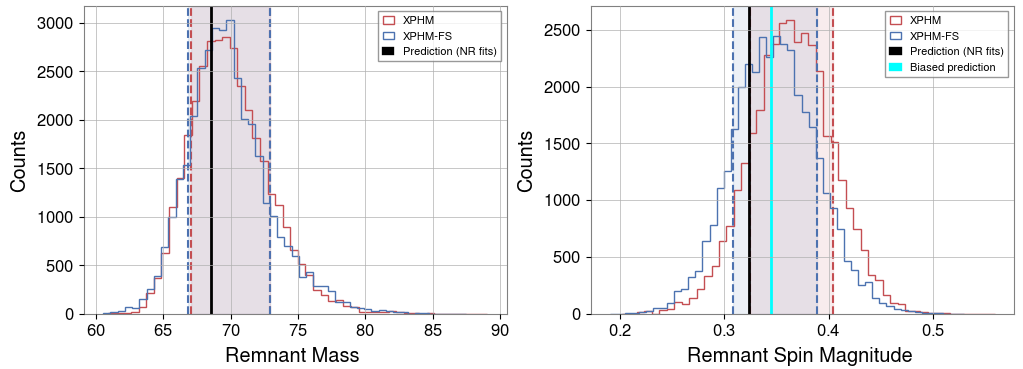}
      \captionsetup{width=1.0\linewidth}
      \caption{Lower SNR injection, with SNR = 20.}
      \label{fig:injectxphm20}
  \end{subfigure}
  \begin{subfigure}[t]{1.0\textwidth}
      \centering
      \includegraphics[width = 0.93\textwidth]{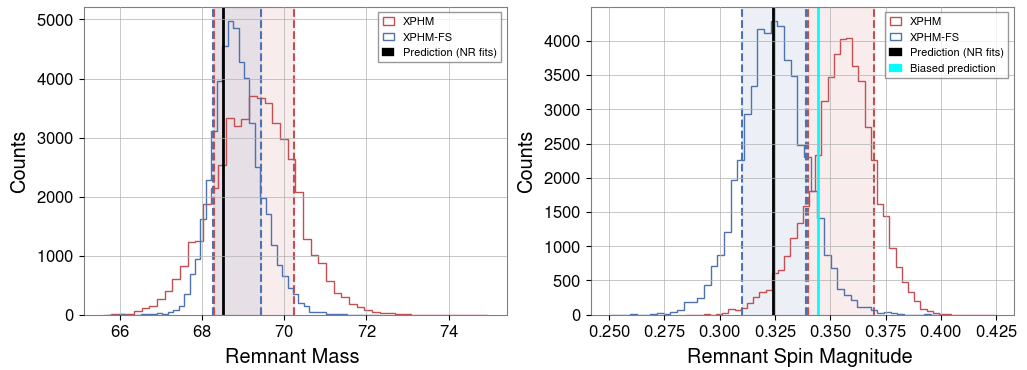}
      \captionsetup{width=1.0\linewidth}
      \caption{Higher SNR injection, with SNR = 60.}
      \label{fig:injectxphm60}
  \end{subfigure}
  \caption{Injection and parameter estimation study results at two different SNR values, for a system with $M=70, q=3, \chi=0.8$ and $\theta = 150\degree$. For both figures, the remnant mass is in units [$M_\odot$]. The predicted values are remnant parameters estimated from the two sections of the waveform, following analysis detailed in Sec. \ref{inspiralremnantparam} and Sec. \ref{ringdownparamestimate}. The shaded regions show the $1\sigma$ confidence regions.}
  \label{fig:injectxphm}
\end{figure*}

In this section, we wish to investigate the effect of the systematic bias due to the mismodelling of ringdown frequency on a complete parameter estimation analysis.
We seek to understand how parameter estimation may be biased, in particular for high mass systems where the merger and ringdown lie in the most sensitive part of the detector.
We inject waveforms and run Bayesian analysis using parameter estimation pipelines.
Attempts were also made to conduct IMR-consistency test injections and parameter estimation; however, the parameter recovery for the ringdown section was too poor to isolate the effects of the systematic bias under investigation.

Our injected waveform uses a modified version of IMR\textsc{Phenom}XPHM. IMR\textsc{Phenom}XPHM is built using the aligned-spin model IMR\textsc{Phenom}XHM \cite{Garcia-Quiros:2020qpx} in the co-precessing frame, before applying a twisting-up procedure. In our analysis here, we use a version of IMR\textsc{Phenom}XPHM with an altered co-precessing frame model called XHM-CP \cite{Hamilton:2021pkf,Thompson:2023ase}. 
This modified version of the aligned spin model IMR\textsc{Phenom}XHM was used as part of the co-precessing frame model for the new precessing waveform model IMR\textsc{Phenom}XO4a \cite{Thompson:2023ase}. 
XHM-CP shifts the QNM frequencies in the co-precessing frame, such that they are then accurately modelled in the inertial frame. 
Other small improvements are included in XHM-CP to improve the accuracy of the model as a whole, mainly affecting the late inspiral-merger section of the model. 
The dominant change in this co-precessing frame model is the ringdown frequencies, and we thus expect our analysis here to reflect specifically the effect of mismodelling the co-precessing ringdown frequency.
We call this version of IMR\textsc{Phenom}XPHM with the correct shifted QNM frequency XPHM-FS. 

We inject waveforms using XPHM-FS, and then recover parameters using both XPHM-FS and XPHM, models—the latter still containing the mismodelled QNM frequencies. 
This parameter estimation is done on the full waveform, instead of a waveform split in two sections for an IMR-consistency test.
Given the computational intensity of this analysis, it was conducted on a limited number of configurations. 
In total, we performed injections for eight sets of intrinsic parameters $(q,\chi,\theta)$, which were $\{ (3, 0.2, 0\degree), (3, 0.2, 60\degree), (3, 0.2, 90\degree), (3, 0.8, 90\degree), \\ (3, 0.8, 150\degree), (8, 0.2, 60\degree), (8, 0.8, 90\degree), (8, 0.8, 150\degree)\}$. 
For each of these parameters, we injected at total mass $70M_\odot$ and $100M_\odot$, and at SNRs 20 and 60. 
The full set of parameters used for the injection analysis shown in Fig. \ref{fig:injectxphm} is listed in Table \ref{tab:allinputparams}. 
The extrinsic parameters were constant for all configurations considered. 
For each of the injections, we considered a three detector network, LIGO Livingston, LIGO Hanford, and VIRGO. The corresponding advanced design sensitivity curve were used for each of these detectors \cite{lvknetwork, LIGOScientific:2014pky,VIRGO:2014yos}. 

The choice of total mass $70M_\odot$ means that the ringdown frequency of the waveform is close to 100 - 200 Hz, which is the most sensitive frequency band of the LIGO and VIRGO detectors \cite{lvknetwork, LIGOScientific:2014pky,VIRGO:2014yos}. 
For systems with a high total mass such as this, the ringdown section of the signal will dominate our data, and mismodelling in the ringdown section will then greatly affect our parameter estimation results. 

Right ascension, declination, polarisation, and inclination were chosen to ensure that the power in the plus and cross polarisations is approximately equal, enhancing the measurability of precession. 
Specifically, our goal was to achieve a ratio close to 1, expressed as $\sqrt{\sum_d{F^2_{+,d}/\sum_dF^2_{\times,d}}}\sim 1$, where $d$ represents the detectors. 
The choice of inclination was also optimised to maximise the observability of precession \cite{Fairhurst:2019vut, Green:2020ptm}.

We then performed Bayesian inference via the \texttt{Bilby} parameter estimation software package \cite{Ashton:2018jfp,10.1093/mnras/staa2850} employing the \texttt{Dynesty} nested sampler \cite{Speagle_2020} with 1000 live points and broad, uninformative priors. 
The injections and parameter estimation were performed using tools from \texttt{yumeng.xu/GWUtils} \cite{gwutils} on the LIGO \texttt{Hawk} computing cluster. 
The results were then processed and visualized using the \texttt{pesummary} package \cite{Hoy:2020vys}, which also facilitated the conversion of inspiral parameter posteriors to remnant parameter posteriors.

\begin{table}[h]
  \centering
  \begin{tabular}{|c|c|}
      \hline
      Total Mass $M$[$M_\odot$] & 70 \\
      Mass Ratio $q$ & 3.0 \\
      Spin Magnitude $\chi$ & 0.8 \\
      \ \ \ \ \ Spin Inclination $\theta$[\degree] \ \ \ \ \ & 150\\
      \hline  
      Right ascension $\alpha$ &  0.254  \\
      Declination $\delta$ & -0.111  \\
      Polarisation $\Psi$ &  4.697 \\
      Inclination $\iota$ & 0.6 \\
      \hline
  \end{tabular}
  \caption{Parameters of the injection for results shown in Fig. \ref{fig:injectxphm}. }
  \label{tab:allinputparams}
\end{table}

Our injections range from aligned-spin to highly precessing systems, and in particular systems with high spin magnitudes and $\theta \approx 150\degree$. The results of our parameter estimation are largely aligned with our expectations. For the $q=3$ systems, the aligned spin and close to aligned spin systems showed good agreement between remnant parameter posteriors estimated from inspiral and merger-ringdown. As the spin magnitude $\chi$ and angle $\theta$ increase, discrepancies in the estimated remnant parameters become more pronounced, reaching a peak in the system outlined in Table \ref{tab:allinputparams}. 
Systems with $q=8$ also exhibited some discrepancies in the estimated remnant parameters, exceeding the anticipated levels due to inaccuracies in modelling the ringdown frequency. This suggests that other systematics, such as the mismodelling of precessional dynamics through merger and ringdown, dominate the results such that we are unable to identify the impact of a given sources of systematic bias. 

The findings of the analysis for the system described in Table \ref{tab:allinputparams} are illustrated in Fig. \ref{fig:injectxphm}, which shows parameter estimation results at SNRs of 20 and 60. The remnant mass posteriors for the two models agree at 1$\sigma$, further justifying our assumption that $M_f^\textrm{insp} = M_f^\textrm{merg}$. 
In this figure, the `Prediction (NR fits)' line on the right correspond to the true injected value calculated from NR fits, while the `Biased prediction' line corresponds to the wrong value that we expect the mismodelling of the ringdown frequency to give.
At lower SNR, shown in Fig. \ref{fig:injectxphm20}, we see that both models recover the true injected value to 1$\sigma$ confidence. 
However, at a higher SNR of 60, depicted in Fig. \ref{fig:injectxphm60}, the disparity between the two posteriors for the final spin becomes evident, with XPHM failing to accurately recover the true injected value within the $1\sigma$ range, leaning instead towards the biased predicted value. 
This highlights the systematic bias that will occur when performing parameter estimation with any waveform model with this mismodelling of the ringdown frequency. 

The majority of events detected thus far have shown no clear evidence of precession and are at lower SNRs, belonging to the region of parameter space where we do not see biases from mismodelling this ringdown frequency. In the future, we expect the effects of this mismodelling to be relevant as detectors become increasingly sensitive and we begin to see unambiguously precessing signals.

\section{\label{5conclusion}Conclusion}
With the ongoing observing run of the LVK \cite{O4announcement}, we anticipate the detection of a significant number of GW events with higher SNRs than previously seen. 
Additionally, we expect to observe signals from systems with considerable precession. It is crucial to pinpoint the limitations of our current waveform models to properly interpret this influx of data.

In this work, we explored a source of systematic bias stemming from the inaccurate calculation of the co-precessing frame QNM frequency $\omega^\textrm{CP}_{22}$, which results in an erroneous final inertial frame QNM frequency $\omega^J_{22}$, and consequently, incorrect phenomenology in the post-merger waveform. 
We considered IMR\textsc{Phenom}XPHM, a highly computationally efficient waveform model currently being used for data analysis.
We identified the regions of the parameter space where this bias is strongest: namely where $(q,\chi)$ are high, and $\theta\sim 150 \degree$. Though the exact points of parameter space where this mismodelling is relevant is waveform model dependent, we expect the general picture to be model independent. 

We examined the bias in the context of the IMR-consistency test by using the Fisher matrix formalism, checking whether the predicted remnant spin $\chi_f$ from the waveform model itself was consistent between inspiral and merger-ringdown portions of the waveform. 
The computational efficiency of the Fisher matrix formalism enabled us to run this analysis across the entire parameter space. Our results demonstrated discrepancies between the inspiral and merger-ringdown sections of the model, for high values of $(q, \chi)$, and $\theta\sim 150\degree$ as expected. 
We found that the discrepancies are greater at larger SNR.

Additionally, we found that employing models that do not correctly account for the effect of precession in the ringdown frequency can lead to biased parameter estimation results, which is particlarly significant at higher SNRs. 

In conclusion, our results have shown how failing to correctly account for all physical effects can lead to waveform systematics that negatively impact both parameter estimation and tests of GR. Here we have focused on precession effects in the merger-ringdown portion of the signal. The majority of currently available waveform models for use in analysis do not completely model precession through merger and ringdown accurately. In addition, we are yet to have full GR-informed models that account for both precession and eccentricity, which will add further biases in parameter estimation.

\section{Acknowledgements}

The authors of this paper would like to thank Shubhanshu Tiwari, Yumeng Xu, Philippe Jetzer, Héctor Estellés, Peter James Nee, Nihar Gupte, Raj Patil, and Elisa Maggio for useful discussions. We would also like to thank Krishnendu N V for her useful comments on the manuscript.

EH was supported in part by Swiss National Science Foundation (SNSF) grant IZCOZ0-189876 and by  UZH Postdoc Grant (Forschungskredit) FK-22-115. EH was also supported in part by the Universitat de les Illes Balears (UIB); the Spanish Agencia Estatal de Investigaci\'{o}n grants PID2022-138626NB-I00, PID2019-106416GB-I00, RED2022-134204-E, RED2022-134411-T, funded by MCIN/AEI/10.13039/501100011033; the MCIN with funding from the European Union NextGenerationEU/PRTR (PRTR-C17.I1); Comunitat Auton\`{o}ma de les Illes Balears through the Direcci\'{o} General de Recerca, Innovació I Transformaci\'{o} Digital with funds from the Tourist Stay Tax Law (PDR2020/11 - ITS2017-006), the Conselleria d’Economia, Hisenda i Innovaci\'{o} grant numbers SINCO2022/18146 and SINCO2022/6719, co-financed by the European Union and FEDER Operational Program 2021-2027 of the Balearic Islands; the ``ERDF A way of making Europe''.

The authors are grateful for computational resources provided by the Cardiff University and supported by STFC grants ST/I006285/1 and ST/V005618/1.

\newpage
\bibliography{biblio} 

\end{document}